\documentclass[aps,prd,groupedaddress,showpacs,superscriptaddress,twocolumn,eqsecnum,floatfix]{revtex4-1}
\usepackage{graphicx}% Include figure files
\usepackage{dcolumn}% Align table columns on decimal point
\usepackage{bm}% bold math
\begin{document}

\title {Dynamical analysis of the $X$ resonance contributions to the decay $J/\psi\to\gamma X\to\gamma\phi\phi$.}

\author{A.~A.~Kozhevnikov}
\email[]{kozhev@math.nsc.ru} \affiliation{Laboratory of
Theoretical Physics, S.~L.~Sobolev Institute for Mathematics, 630090, Novosibirsk, Russian
Federation}
\affiliation{Novosibirsk State University, 630090, Novosibirsk, Russian
Federation}

\date{\today}

\begin{abstract}
The dynamics of the  $J^{PC}=0^{-+}$, $0^{++}$, and $2^{++}$ resonance contributions  to the decay $J/\psi\to\gamma X(J^{PC})\to\gamma\phi\phi$ is analysed using the data obtained by BESIII collaboration. The effective coupling constants parameterising invariant amplitudes of the transitions $J/\psi\to\gamma X(J^{PC})$ and $X(J^{PC})\to\phi\phi$ and masses of $X(J^{PC})$ resonances are found from the fits. They are used for evaluation of the branching fractions $B_{X(J^{PC})\to\phi\phi}$, relative branching fractions $B_{J/\psi\to\gamma X(J^{PC})\to\gamma\phi\phi}$, and for  obtaining the photon angular distributions.
\end{abstract}
%\pacs{13.25.Hw,14.40.Cs,14.40.Nd}

\maketitle

\section{Introduction}
\label{intro}

The interest in the decay $J/\psi\to\gamma\phi\phi$ \cite{bisello1986,bai1990,ablikim2008,BES16} is related  with the possible  existence of the exotic glueball state  decaying into the $\phi\phi$ pair \cite{Linden,Etkin85,Booth86,Etkin88}. The spin-parity quantum numbers of the resonance states decaying into $\phi\phi$ are reported to be $J^P=0^+ , 0^-$, and $2^+$ \cite{ablikim2008,BES16,PDG}.

\begin{figure}
\includegraphics[width=5cm]{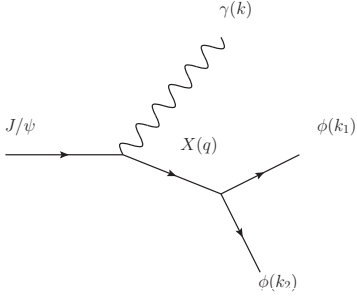}
\caption{\label{diag}The dominant diagram of the decay
$J/\psi\to\gamma\phi\phi$.}\end{figure}

The partial wave analysis of the $\phi\phi$ system was performed in Ref.~\cite{BES16} based on the model  with the coherent sum of the Breit-Wigner amplitudes with the constant widths,
\begin{equation}\label{simpleBW}
M_X\propto\frac{1}{m^2_X-s-im_X\Gamma_X}.\end{equation}
As is pointed out in Ref.~\cite{BES16}, the process $J/\psi\to f_1(1285)\phi\to\gamma\phi\phi$ can be neglected because of its small branching fraction,
hence the diagram for the decay $J/\psi\to\gamma\phi\phi$  shown in Fig.~\ref{diag} is assumed  to be dominant.
%--------------------------------------------------------------------------------
%--------------------------------------------------------------------------------

However, the dynamics of the decay chain $J/\psi\to\gamma X(J^P)$, $X(J^P)\to\phi\phi$  is relatively simple only in case of the pseudoscalar resonance admitting the single contribution with the unit orbital momentum in both above vertices. In general, one should include the  different spin-orbital momentum  structures for different spin-parities of the $X(J^{P})$ resonances in the $\phi\phi$ system, especially in case of the tensor contribution $J^{PC}=2^{++}$ where a number of independent spin structures enter the amplitudes of transitions $J/\psi\to\gamma X(2^+)$ and $X(2^+)\to\phi\phi$. So it is reasonable to reanalyze the data of Ref.~\cite{BES16} in the model with the energy-dependent partial $\phi\phi$ width in order to extract the magnitudes of the effective coupling constants parametrizing the effective invariant amplitudes of the above transitions. This is the goal of the present work. The data \cite{BES16} will be described here by taking, as the starting point, the effective amplitudes in the invariant form parametrized by some unknown constants. Then their three-dimensional counterparts will be written which take into account the condition of the three-dimensional transverse character of the final photon polarization vector. The three-dimensional form simplifies considerably the derivation of expressions for the $\phi\phi$ mass spectrum and angular distributions.

The kinematic notations are the following. The four-momenta assignment is $J/\psi(Q)\to\gamma(k)X(q)\to\gamma(k)\phi(k_1)\phi(k_2)$; $\epsilon_\mu$,  $\epsilon_{1\mu}$, $\epsilon_{2\mu}$ (${\bm\xi}, {\bm\xi}_1,{\bm\xi}_2$) are, respectively, the polarization four-vectors of the $J/\psi$ meson  and $\phi$ mesons (their 3-dimensional counterparts in their respective rest frame);  $e_\mu=(0,{\bm e})$ stands for the polarization four-vector of the photon, $\epsilon_{\mu\nu\lambda\sigma}$ is the Levy-Civita tensor. The energy-momentum 4-vector of the $\phi\phi$ state in the $J/\psi$ rest frame is $q=(q_0,{\bm q})$,
\begin{eqnarray}\label{qmu}
q_0&=&\frac{m^2_{J/\psi}+m^2_{12}}{2m_{J/\psi}},\nonumber\\
{\bm q}&=&-{\bm k}=-{\bm n}\frac{m^2_{J/\psi}-m^2_{12}}{2m_{J/\psi}},\end{eqnarray}where ${\bm n}$ stands for the unit vector in the direction of the photon, and $m_{12}$ is the invariant mass of the $\phi\phi$ pair. In turn, the energy-momentum of one of  the $\phi$ mesons, $k_{1\mu}=(k^\ast_{10},{\bm k}^\ast_1)$, in the center-of-mass system of the $\phi\phi$ pair, is
\begin{eqnarray}\label{k1mu}
k^\ast_{10}&=&\frac{1}{2}m_{12},\nonumber\\
{\bm k}^\ast_1&=&\frac{{\bm n}_1}{2}\sqrt{m^2_{12}-4m^2_\phi},
\end{eqnarray}with ${\bm n}_1$ being the unit vector in the direction of the motion of the $\phi$ meson.

The subsequent material is organized as follows. In Sec.~\ref{sec2}, the parametrizations of invariant amplitudes and their three-dimensional counterparts are given, together with the expressions for partial widths. Section \ref{sec3} is devoted to presenting the results of fitting of the different partial wave contributions to the $\phi\phi$ mass spectrum. The discussion presented in Sec.~\ref{sec4} concerns the consistency of the fits, together with the concluding remarks. The $J/\psi\to\gamma X(J^P)\to\gamma\phi\phi$ amplitudes in terms of the independent helicity amplitudes in the $J/\psi\to\gamma X(J^P)$ vertex are given in the Appendix.

\section{Amplitudes and partial widths}\label{sec2}
~

First, let us write the amplitude of the decay $J/\psi\to\gamma X\to\gamma\phi\phi$ assuming, for a while, the single intermediate resonance $X$.
Schematically, the method of evaluation of the amplitudes adopted in the present work is as follows. In the case of the $X$ resonance with spin zero and two one has, respectively
\begin{widetext}
\begin{eqnarray}\label{amps}
M_{J_X=0}(J/\psi\to\gamma X\to\gamma\phi\phi)&=&\frac{[M(J/\psi\to\gamma X)][M(X\to\phi\phi)]}{D_X},\nonumber\\
M_{J_X=2}(J/\psi\to\gamma X\to\gamma\phi\phi)&=&\sum_{\lambda_X}\frac{[M_{\mu\nu}(J/\psi\to\gamma X)T_{\mu\nu}^{(\lambda_X)}]
[M_{\alpha\beta}(X\to\phi\phi)T_{\alpha\beta}^{(\lambda_X)}]}{D_X},
\end{eqnarray}
\end{widetext}
where $D_X$ stands for the inverse propagator of the $X$ resonance. See Eq.~(\ref{prop}) below.
The polarization tensor $T_{\mu\nu}\equiv T_{\mu\nu}^{(\lambda_X)}$ of the spin two resonance is represented in the form
\begin{eqnarray}\label{tij}
T_{00}&=&\frac{t_{ij}q_iq_j}{m^2_{12}},\nonumber\\
T_{0i}&=&\frac{q_j}{m_{12}}\left[t_{ij}+\frac{t_{jk}q_iq_k}{m_{12}(q_0+m_{12})}\right],\nonumber\\
T_{ij}&=&t_{ij}+\frac{(t_{ik}q_j+t_{jk}q_i)q_k}{m_{12}(q_0+m_{12})}+\frac{t_{kl}q_iq_jq_kq_l}{m^2_{12}(q_0+m_{12})^2},
\end{eqnarray}
where $t_{ij}\equiv t_{ij}^{(\lambda_X)}$ is the polarization tensor in the rest frame, so that all the amplitudes can be expressed through the polarization structures in the $X$ rest frame. Since each of the amplitudes in square brackets in Eq.~(\ref{amps}) is Lorentz-invariant one can evaluate it in the respective rest frame, $J/\psi$ or $X$.

\subsection{$J^{PC}=0^{-+}$}\label{0mi}
~

The effective amplitudes for the processes $J/\psi\to\gamma X(0^-)$ and $X(0^-)\to\phi\phi$ and their three-dimensional form in the respective rest frame systems are chosen as follows:
\begin{eqnarray}\label{ampga0mi}
M_{J/\psi\to\gamma X(0^-)}&=&g_{J/\psi\gamma X(0^-)}\epsilon_{\mu\nu\lambda\sigma}Q_\mu\epsilon_\nu k_\lambda e_\sigma=\nonumber\\&&
g_{J/\psi\gamma X(0^-)}m_{J/\psi}|{\bm k}|({\bm n}\cdot[{\bm\xi}\times{\bm e}]),
\end{eqnarray}and
\begin{eqnarray}\label{amp0mifi}
M_{X(0^-)\to\phi\phi}&=&g_{X(0^-)\to\phi\phi}\epsilon_{\mu\nu\lambda\sigma}k_{1\mu}\epsilon_{1\nu}k_{2\lambda}\epsilon_{2\sigma}=\nonumber\\&&
g_{X(0^-)\phi\phi}m_{12}|{\bm k}^\ast_1|({\bm n}_1\cdot[{\bm\xi}_1\times{\bm\xi}_2]).
\end{eqnarray}The calculated partial widths read, respectively,
\begin{equation}\label{Gam0mig}
\Gamma_{J/\psi\to\gamma X(0^-)}(m_{12})=\frac{g^2_{J/\psi\gamma X(0^-)}}{12\pi}|{\bm k}|^3
\end{equation}and
\begin{equation}\label{Gam0mi2fi}
\Gamma_{X(0^-)\to\phi\phi}(m_{12})=\frac{g^2_{X(0^-)\phi\phi}}{8\pi}|{\bm k}^\ast_1|^3.
\end{equation}

The amplitude $M_{J/\psi\to\gamma X(0^-)\to\gamma\phi\phi}\equiv M$ is written in the form
\begin{eqnarray}\label{afin0mi}
M&=&A^{(0^-)}m_{J/\psi}m_{12}|{\bm k}||{\bm k}^\ast_1|\times\nonumber\\&&
({\bm\xi}[{\bm n}\times{\bm e}])({\bm n}_1[{\bm\xi}_1\times{\bm\xi}_2]).
\end{eqnarray}The dynamics of process is included through the factor $A^{(0^-)}$ to be specified below.
The modulus squared summed over polarizations of final particles but keeping the $J/\psi$ polarization fixed reads
\begin{eqnarray}\label{M2sum0mi}
\sum_{\lambda_\gamma\lambda_1\lambda_2}|M|^2&=&2m^2_{J/\psi}m^2_{12}{\bm k}^2{\bm k}^{\ast2}_1\left|A^{(0^-)}\right|^2\times\nonumber\\&&\left[{\bm\xi}\times{\bm n}\right]^2.
\end{eqnarray}Since $\int \left[{\bm\xi}\times{\bm n}\right]^2d\Omega_{\bm n}/4\pi=2{\bm\xi}^2/3$, the decay rate integrated over photon direction is the same for all $J/\psi$ projections.

\subsection{$J^{PC}=0^{++}$}\label{0pl}
~

Since the photon polarization four-vector is $e_\mu=(0,{\bm e})$, the D-wave structure in the $J/\psi\to\gamma X(0^+)$ amplitude vanishes, hence
\begin{equation}\label{Mga0pl}
M_{J/\psi\to\gamma X(0^+)}=-g_1(\epsilon e)=g_1({\bm\xi}{\bm e}).
\end{equation}Correspondingly, the invariant amplitude of the decay $X(0^+)\to\phi\phi$ and its three-dimensional form in the $X$ rest frame look as follows:
\begin{eqnarray}\label{Msc2fi}
M_{X(0^+)\to\phi\phi}&=&-f_1(\epsilon_1\epsilon_2)-f_2(\epsilon_1k_2)(\epsilon_2k_1)=\nonumber\\&&f^{(0^+)}_{00}({\bm\xi}_1{\bm\xi}_2)+
f^{(0^+)}_{22}({\bm\xi}_1{\bm n}_1)\times\nonumber\\&&({\bm\xi}_2{\bm n}_1),
\end{eqnarray}where
\begin{eqnarray}\label{fLSsc}
f^{(0^+)}_{00}&=&f_1,\nonumber\\
f^{(0^+)}_{22}&=&(2f_1+f_2m^2_{12})\frac{{\bm k}^{\ast2}_1}{m^2_\phi}.
\end{eqnarray}As is evident from these expressions, $f^{(0^+)}_{00}$ and $f^{(0^+)}_{22}$ correspond to the assignment $(S,L)=(0,0)$ and $(2,2)$, respectively, of the spin $S$ and orbital angular momentum $L$ of the $\phi\phi$ state. The energy-dependent partial widths look like
\begin{eqnarray}\label{widthsc}
\Gamma_{J/\psi\to\gamma X(0^+)}&=&\frac{g^2_1|{\bm k}|}{12\pi m^2_{J/\psi}},\nonumber\\
\Gamma_{X(0^+)\to\phi\phi}&=&\frac{|{\bm k}^\ast_1|}{16\pi m^2_{12}}\times\nonumber\\&&(2|f^{(0^+)}_{00}|^2+|f^{(0^+)}_{00}+f^{(0^+)}_{22}|^2).
\end{eqnarray}The dynamical content of the $J/\psi\to\gamma X(0^+)\to\gamma\phi\phi$ component of the $\phi\phi$ spectrum will be specified below in subsection \ref{subsec3b}.

\subsection{$J^{PC}=2^{++}$}\label{2pl}
~

The invariant amplitude of the transition $J/\psi\to\gamma X(2^+)$ and its three-dimensional form  are the following:
\begin{widetext}
\begin{eqnarray}
M_{J/\psi\to\gamma X(2^+)}&=&\left[c_1(\epsilon e)Q_\mu Q_\nu+c_2(\epsilon k)e_\mu k_\nu+c_3\epsilon_\mu e_\nu\right]T_{\mu\nu}\equiv\left[g_{02}({\bm\xi}\cdot{\bm e})n_in_j+g_{12}({\bm\xi}\cdot{\bm n})e_in_j+g_{20}\xi_ie_j\right]t_{ij},
\label{Mga2pl}
\end{eqnarray}
\end{widetext}
where $c_{1,2,3}$ are, in principle, the functions of the invariant mass $m_{12}$. For nothing better, we assume them to be some, in general, complex constants;
\begin{eqnarray}
g_{02}&=&-c_1\frac{m^2_{J/\psi}{\bm k}^2}{m^2_{12}},\nonumber\\
g_{12}&=&-\frac{{\bm k}^2}{m_{12}}\left(c_2q_0+\frac{c_3}{q_0+m_{12}}\right),\nonumber\\
g_{20}&=&-c_3,
\label{gLS}
\end{eqnarray}where ${\bm k}=-{\bm q}$. See $(q_0,{\bm q})$ in Eq.~(\ref{qmu}).

In turn, the invariant amplitude of the decay $X\to\phi\phi$ and its three-dimensional presentation in the $\phi\phi$ center-of-mass system look as follows:
\begin{widetext}
\begin{eqnarray}\label{MX2pl}
M_{X(2^+)\to\phi\phi}&=&\left\{g_1\epsilon_{1\mu}\epsilon_{2\nu}+k_{1\mu}k_{2\nu}\left[g_2(\epsilon_1\epsilon_2)+
g_3(\epsilon_1k_2)(\epsilon_2k_1)\right]+g_4\left[\epsilon_{1\mu}k_{2\nu}(\epsilon_2k_1)+
\epsilon_{2\mu}k_{1\nu}(\epsilon_1k_2)\right]\right\}T_{\mu\nu}\equiv\nonumber\\&& \left[f_{20}\xi_{1i}\xi_{2j}+f_{02}({\bm\xi}_1\cdot{\bm\xi}_2)n_{1i}n_{1j}+
f_{22}\left[({\bm\xi}_1\cdot{\bm n}_1)\xi_{2i}+({\bm\xi}_2\cdot{\bm n}_1)\xi_{1i}\right]n_{1j}+\right.\nonumber\\&&\left.f_{24}({\bm\xi}_1\cdot{\bm n}_1)({\bm\xi}_2\cdot{\bm n}_1)n_{1i}n_{1j}\right]t_{ij}.
\end{eqnarray}
\end{widetext}
Here,
\begin{eqnarray}\label{fLS}
f_{20}&=&g_1,\nonumber\\
f_{02}&=&g_2{\bm k}_1^{\ast2},\nonumber\\
f_{22}&=&\frac{{\bm k}_1^{\ast2}}{m_\phi}\left(\frac{g_1}{k^\ast_{10}+m_\phi}+g_4m_{12}\right),\nonumber\\
f_{24}&=&\frac{{\bm k}_1^{\ast4}}{m_\phi^2}\left[\frac{g_1}{(k^\ast_{10}+m_\phi)^2}+2g_2+g_3m^2_{12}+\right.\nonumber\\&&\left.
2g_4\frac{m_{12}}{k^\ast_{10}+m_\phi}\right].\end{eqnarray}Again, the indices at the quantities in the left-hand side of these equations refer to the possible spin-orbital momentum assignments $(S,L)=(2,0)$, $(0,2)$, $(2,2)$, and $(2,4)$ of the $\phi\phi$ state.

The sum over polarizations of the intermediate tensor resonance is fulfilled with the help of relation
\begin{eqnarray}\label{polsum2}
\sum_{\lambda_X}t^{(\lambda_X)}_{ij}t^{(\lambda_X)}_{kl}&=&\frac{1}{2}(\delta_{ik}\delta_{jl}+\delta_{il}\delta_{jk})-
\frac{1}{3}\delta_{ij}\delta_{kl}\nonumber\\&&\equiv\Pi_{ij,kl}.
\end{eqnarray}

The modulus squared of the amplitude $M\equiv M_{J/\psi\to\gamma X(2^+)\to\gamma\phi\phi}$ summed over polarizations of final particles can be represented in the following form suitable for subsequent integrations over final states:
\begin{eqnarray}\label{sumsq2}
\sum_{\lambda_\gamma\lambda_1\lambda_2}|M|^2&=&I_{ij,c}I^\ast_{i^\prime j^\prime,c^\prime}(\delta_{cc^\prime}-n_cn_{c^\prime})\times\nonumber
\\&&\frac{\Pi_{ij,kl}\Pi_{i^\prime j^\prime,k^\prime l^\prime}}{|D_{X(2^+)}|^2}F_{kl,ab}F^\ast_{k^\prime l^\prime,ab},
\end{eqnarray}where $D_{X(2^+)}\equiv D_{X(2^+)}(m^2_{12})$ is given by Eq.(\ref{prop}),
\begin{eqnarray}\label{IF}
I_{ij,c}&=&g_{02}n_in_j\xi_c+\left[g_{12}({\bm\xi}{\bm n})n_j+g_{20}\xi_j\right]\delta_{ic},\nonumber\\
F_{kl,ab}&=&f_{20}\delta_{ka}\delta_{lb}+f_{02}\delta_{ab}n_{1k}n_{1l}+\nonumber\\&&f_{22}(n_{1a}\delta_{kb}+n_{1b}\delta_{ka})n_{1l}+
f_{24}n_{1k}n_{1l}\times\nonumber\\&&n_{1a}n_{1b}.
\end{eqnarray}The $\phi\phi$ mass spectrum in the decay $J/\psi\to\gamma X(2^+)\to\gamma\phi\phi$ can be written as
\begin{widetext}
\begin{eqnarray}\label{spec2pl}
\frac{d\Gamma}{dm_{12}}&=&\frac{|{\bm k}||{\bm k}_1^\ast|}{32\pi^3m^2_{J/\psi}}\left[\int I_{ij,c}I^\ast_{i^\prime j^\prime,c^\prime}(\delta_{cc^\prime}-n_cn_{c^\prime})\frac{d\Omega_{{\bm n}}}{4\pi}\right]\left[\int F_{kl,ab}F^\ast_{k^\prime l^\prime,ab}\frac{d\Omega_{{\bm n}_1}}{4\pi}\right]\frac{\Pi_{ij,kl}\Pi_{i^\prime j^\prime,k^\prime l^\prime}}{|D_{X(2^+)}|^2}.
\end{eqnarray}
\end{widetext}Note that the polarization state of the $J/\psi$ meson is kept fixed for a while. Since in terms of quantities designated by square brackets in Eq.~(\ref{spec2pl}) the partial widths of the decays $J/\psi\to\gamma X(2^+)$, $X(2^+)\to\phi\phi$ look, respectively, like
\begin{widetext}
\begin{eqnarray}\label{Gampsi2}
 \Gamma_{J/\psi\to\gamma X(2^+)}&=&\frac{|{\bm k}|}{8\pi m^2_{J/\psi}}\left[\int I_{ij,c}I^\ast_{i^\prime j^\prime,c^\prime}(\delta_{cc^\prime}-n_cn_{c^\prime})\frac{d\Omega_{{\bm n}}}{4\pi}\right]\Pi_{ij,i^\prime j^\prime}=\frac{|{\bm k}|}{8\pi m^2_{J/\psi}}\int\left\{\left(|g_{02}|^2+\frac{3}{2}|g_{20}|^2-\right.\right.\nonumber\\&&\left.\left.
 \frac{1}{3}|g_{02}+g_{20}|^2\right)[{\bm\xi}\times{\bm n}]^2+|g_{12}+g_{20}|^2({\bm\xi}{\bm n})^2\right\}\frac{d\Omega_{{\bm n}}}{4\pi},
\end{eqnarray}
(polarization of $J/\psi$ is still fixed)
\begin{eqnarray}\label{GamX2}
\Gamma_{X(2^+)\to\phi\phi}&=&\frac{|{\bm k}^\ast_1|}{80\pi m^2_{12}}\left[\int F_{kl,ab}F^\ast_{k^\prime l^\prime,ab}\frac{d\Omega_{{\bm n}_1}}{4\pi}\right]\Pi_{kl,k^\prime l^\prime}=\frac{|{\bm k}^\ast_1|}{240\pi m^2_{12}}\left(10|f_{20}+f_{22}|^2+3|f_{20}|^2+2|f_{20}+f_{24}|^2+\right.\nonumber\\&&\left.2|f_{02}+f_{24}|^2+4|f_{02}+f_{22}|^2+
4|f_{22}+f_{24}|^2-4|f_{22}|^2-6|f_{24}|^2\right),
\end{eqnarray}
\end{widetext}
then the $\phi\phi$ mass spectrum in the decay $J/\psi\to\gamma X(2^+)\to\gamma\phi\phi$ in case of the single intermediate resonance can be written in the standard form:
\begin{eqnarray}\label{spectrum}
\frac{d\Gamma}{dm_{12}}&=&\frac{2m^2_{12}\Gamma_{J/\psi\to\gamma X(2^+)}\Gamma_{X(2^+)\to\phi\phi}}{\pi|D_{X(2^+)}|^2}.
\end{eqnarray}Taking into account of a number of interfering resonances with given spin-parity demands the modifications analogous to those discussed in subsections \ref{subsec3a} and \ref{subsec3b}.

\section{Results}
\label{sec3}
~

When fitting the data Ref.~\cite{BES16}, the coupling constants characterizing invariant amplitudes Eqs.~(\ref{ampga0mi}), (\ref{amp0mifi}),  (\ref{Mga0pl}), (\ref{Msc2fi}), (\ref{Mga2pl}), and (\ref{MX2pl}) are  assumed to be real. Nonzero imaginary parts would point to the  dynamical effects related with the re-scattering of the final $\phi$ mesons \cite{fn1}. The quantitative inclusion of these effects would require the introduction of multiple additional parameters such as coupling constants of the exchanged particles with $\phi$ mesons, the slope parameters characterizing  the above exchange etc. This seems to be premature with  the present accuracy of the data.

\subsection{Pseudoscalar resonance contribution}\label{subsec3a}
~

The expression for the $J^{PC}=0^{-+}$ resonance component of the spectrum  averaged over $\lambda_{J/\psi}=\pm1$ with the help of Eq.~(\ref{M2sum0mi}) is
\begin{eqnarray}\label{spec0mi}
\frac{dN^{(0^-)}}{dm_{12}}&=&\frac{{\cal N}}{(2\pi)^3\times6}\left|A^{(0^-)}\right|^2m^2_{12}|{\bm k}|^3|{\bm k}^\ast_1|^3,
\end{eqnarray}where ${\cal N}$ is unknown overall normalization factor. Three  intermediate resonances  were included in the partial wave analysis of Ref.~\cite{BES16} to describe the partial wave with $J^P=0^-$. We designate  them as $X_1=\eta(2225)$, $X_2=\eta(2100)$, and $X_3=X(2500)$.  Since the case of pseudoscalar resonance $X$ is kinematically simple, we consider two models for the amplitude.

(i) The model A. It allows for inclusion of the mixing of the above  three resonances via their common decay mode $\phi\phi$ in the form used earlier in Ref.~\cite{ach97,ach02}:
\begin{widetext}
\begin{eqnarray}\label{Amix}
A^{(0^-)}&=&\left(
              \begin{array}{ccc}
                g_{J/\psi\gamma X_1} & g_{J/\psi\gamma X_2} & g_{J/\psi\gamma X_3} \\
              \end{array}
            \right)
\left(
  \begin{array}{ccc}
    D_1 & -\Pi_{12} & -\Pi_{13} \\
    -\Pi_{12} & D_2 & -\Pi_{23} \\
    -\Pi_{13} & -\Pi_{23} & D_3 \\
  \end{array}
\right)^{-1}\left(
              \begin{array}{c}
                g_{X_1\phi\phi} \\
                g_{X_2\phi\phi} \\
                g_{X_3\phi\phi} \\
              \end{array}
            \right).
\end{eqnarray}
\end{widetext}Let us specify the elements of the matrix of inverse propagators in Eq.~(\ref{Amix}). The main goal here is to analyze the $\phi\phi$ decay mode of the $X$ resonances. However, another decay modes are feasible. We will assume that $\phi\phi$ is the only common decay mode and take its energy dependence  in $D_i$, $\Pi_{ij}$ explicitly ($i,j=1,2,3$),
while other decay modes will be effectively taken into account in the fixed width  approximation, $\Gamma^\prime_{X_i}=$const. Specifically, the inverse propagator of the resonance $X_i$ is assumed to be
\begin{eqnarray}\label{prop}
D_{X_i(J^P)}(m^2_{12})&=&m^2_{X_i(J^P)}-m^2_{12}-\nonumber\\&&im_{12}\Gamma_{X_i(J^P)\to\phi\phi}(m_{12})-\nonumber\\&&
im_{X_i(J^P)}\Gamma^\prime_{X_i}.\end{eqnarray}The polarization operator $\Pi_{ij}$ responsible for the mixing looks as follows:
\begin{eqnarray}\label{Pi}
\Pi_{ij}&\equiv&\Pi_{ij}(m^2_{12})={\rm Re}\Pi_{ij}+im_{12}g_{X_i\phi\phi}g_{X_j\phi\phi}\times\nonumber\\&&
\frac{|{\bm k}^\ast_1|^3}{8\pi}.
\end{eqnarray}
Here, ${\rm Im}\Pi_{ij}$ is fixed by the unitarity  relation while ${\rm Re}\Pi_{ij}$, in principle, can be evaluated through the dispersion relation. However, taken literally, the dispersion integral is divergent due to the fast growth with energy of the $X(0^-)\to\phi\phi$ partial width, and one should introduce the phenomenological suppression factor parametrized by some unknown constant, in order to make the integral finite. We take here the practical attitude and assume that ${\rm Re}\Pi_{ij}$ are some constants, $a_{12}\equiv{\rm Re}\Pi_{12}$, $a_{13}\equiv{\rm Re}\Pi_{13}$, and $a_{23}\equiv{\rm Re}\Pi_{23}$ to be determined from the fit.

(ii) The model B  corresponds to the vanishing  mixing, by taking  $\Pi_{ij}\equiv0$, that is, $A^{(0^-)}$ is given by the coherent sum of the energy dependent Breit -- Wigner terms.

The fitted parameters are the following: $m_{X_i}$, $\Gamma^\prime_{X_i}$, $g_{J/\psi\gamma X_i}\sqrt{\cal N}$, $g_{X_i\phi\phi}$, $i=1,2,3$, in both models A and B, and three additional parameters $a_{12}$, $a_{13}$, and $a_{23}$ in the model A. The results are presented in the Table \ref{tab1}, together with some branching fractions evaluated with the found parameters. To be specific,  evaluated are the  $X_i\to\phi\phi$ branching fractions:
\begin{equation}\label{Br2fi}
B_{X_i\to\phi\phi}=\frac{2}{\pi}\int_{2m_\phi}^{m_{\rm max}}\frac{m^2_{12}\Gamma_{X_i\to\phi\phi}(m_{12})}{|D_{X_i}|^2}dm_{12}.
\end{equation}The answer depends on upper integration limit $m_{\rm max}$. Since the data of Ref.~\cite{BES16} refer to the interval $2m_\phi<m_{12}<2.7$ GeV, we take here $m_{\rm max}=2.7$ GeV. The branching fraction $B_{J/\psi\to\gamma X_i\to\gamma\phi\phi}$, in general, cannot be represented in the form $B_{J/\psi\to\gamma X_i\to\gamma\phi\phi}=B_{J/\psi\to\gamma X_i}\times B_{X_i\to\phi\phi}$, due to the large widths of the resonance $X$ and (or) to the strong energy dependence of the partial width. The general expression is
\begin{widetext}
\begin{equation}\label{Bfin}
B_{J/\psi\to\gamma X_i\to\gamma\phi\phi}=\frac{2}{\pi\Gamma_{J/\psi}}\int_{2m_\phi}^{m_{J/\psi}}\frac{m^2_{12}\Gamma_{J/\psi\to\gamma X_i}(m_{12})\times \Gamma_{X_i\to\gamma\phi\phi}(m_{12})dm_{12}}{|D_{X_i}(m^2_{12})|^2}.
\end{equation}
\end{widetext}It reduces to the mentioned factorization in the narrow width approximation. Furthermore, because the overall normalization factor ${\cal N}$ is unknown, only the quantities ${\cal N}\Gamma_{J/\psi}B_{J/\psi\to\gamma X_i\to\gamma\phi\phi}$, $i=1,2,3$, and
\begin{eqnarray}\label{Np}
N^{(0^-)}&\equiv&{\cal N}\Gamma_{J/\psi}B_{J/\psi\to\gamma (X_1+X_2+X_3)\to\gamma\phi\phi}=\nonumber\\&&\int_{2m_\phi}^{m_{J/\psi}}\frac{dN^{(0^-)}}{dm_{12}}dm_{12}\end{eqnarray} are presented, not the absolute branching fractions. One can evaluate the role of interference,
\begin{widetext}
\begin{eqnarray}\label{interf}
I&=&{\cal N}\Gamma_{J/\psi}\left[B_{J/\psi\to\gamma (X_1(0^-)+X_2(0^-)+X_3(0^-))\to\gamma\phi\phi}-
\sum_{i=1,2,3}B_{J/\psi\to\gamma X_i(0^-)\to\gamma\phi\phi}\right],
\end{eqnarray}
\end{widetext} to obtain $I=-262\pm33$ (model A) and $I=151\pm19$ (model B).
%%%%%%%%%%%%%%%%%%%%%%%%%%%%%%%%%%%%%%%%%%%%%%%%%%%%%%%%%%%%%%%%%%%%%%%%%
\begin{table*}
\caption{\label{tab1}Results of fitting the pseudoscalar  resonance contribution in the reaction $e^+e^-\to J/\psi\to\gamma X(0^-)\to\gamma\phi\phi$. The quantity $N^{(0^-)}$ is given by Eq.~(\ref{Np}).}
\begin{ruledtabular}
\begin{tabular}{llll}
 parameter&model A&model B \\ \hline
\\
$m_{X_1(0^-)}$ [GeV]&$2.2312\pm0.0015$&$2.252\pm0.002$\\
$\Gamma^\prime_{X_1(0^-)}$ [GeV]&$0.227\pm0.002$&$0.189\pm0.002$\\
$\sqrt{{\cal N}}g_{J/\psi\gamma X_1(0^-)}$[GeV$^{-3/2}$]&$2450\pm15$&$439\pm3$\\
$g_{X_1(0^-)\phi\phi}$ [GeV$^{-1}$]&$0.881\pm0.005$&$3.49\pm0.03$\\
$B_{X_1(0^-)\to\phi\phi}$&$(1.55\pm0.02)\times10^{-2}$&$0.218\pm0.003$\\
${\cal N}\Gamma_{J/\psi}B_{J/\psi\to\gamma X_1(0^-)\to\gamma\phi\phi}$&$659\pm14$&$300\pm6$\\
\\
$m_{X_2(0^-)}$ [GeV]&$2.0757\pm0.0025$&$2.077\pm0.002$\\
$\Gamma^\prime_{X_2(0^-)}$ [GeV]&$0.136\pm0.0050$&$0.118\pm0.005$\\
$\sqrt{{\cal N}}g_{J/\psi\gamma X_2(0^-)}$[GeV$^{-3/2}$]&$3010\pm110$&$2580\pm80$\\
$g_{X_2(0^-)\phi\phi}$ [GeV$^{-1}$]&$0.160\pm0.011$&$0.29\pm0.01$\\
$B_{X_2(0^-)\to\phi\phi}$&$(2.94\pm0.43)\times10^{-4}$&$(1.03\pm0.08)\times10^{-3}$\\
${\cal N}\Gamma_{J/\psi}B_{J/\psi\to\gamma X_2(0^-)\to\gamma\phi\phi}$&$22\pm4$&$59\pm6$\\
\\
$m_{X_3(0^-)}$ [GeV]&$2.6590\pm0.0028$&$2.705\pm0.003$\\
$\Gamma^\prime_{X_3(0^-)}$ [GeV]&$0.51\pm0.01$&$0.34\pm0.01$\\
$\sqrt{{\cal N}}g_{J/\psi\gamma X_3(0^-)}$[GeV$^{-3/2}]$&$802\pm6$&$591\pm6$\\
$g_{X_3(0^-)\phi\phi}$ [GeV$^{-1}$]&$4.81\pm0.15$&$4.99\pm0.08$\\
$B_{X_3(0^-)\to\phi\phi}$&$(9.72\pm0.26)\times10^{-2}$&$0.126\pm0.003$\\
${\cal N}\Gamma_{J/\psi}B_{J/\psi\to\gamma X_3(0^-)\to\gamma\phi\phi}$&$291\pm26$&$198\pm5$\\
$a_{12}$ [GeV$^2$]&$0.128\pm0.003$&-\\
$a_{13}$ [GeV$^2$]&$-0.087\pm0.004$&-\\
$a_{23}$ [GeV$^2$]&$-0.005\pm0.004$&-\\
$N^{(0^-)}$&$710\pm13$&$708\pm16$\\
$\chi^2/n_{\rm d.o.f.}$&$24.4/18\approx1.4$&$25.8/21\approx1.2$\\
\end{tabular}
\end{ruledtabular}
\end{table*}
%%%%%%%%%%%%%%%%%%%%%%%%%%%%%%%%%%%%%%%%%%%%%%%%%%%%%%%%%%%%%%%%%%%%%%%%%%
%--------------------------------------------------------------------------------
\begin{figure}
\includegraphics[width=12cm]{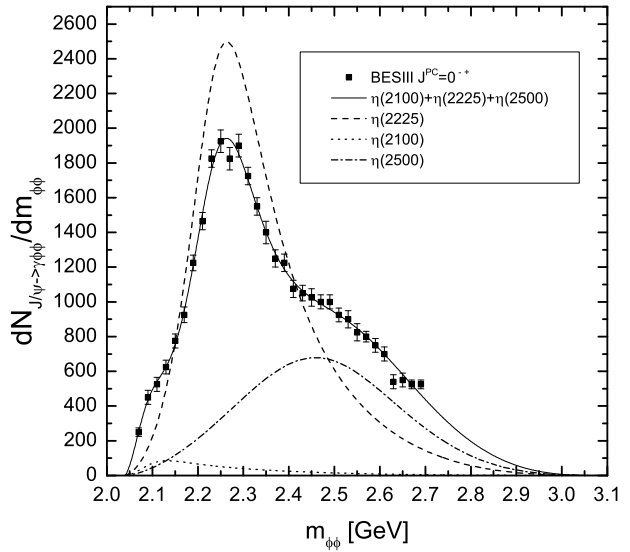}
\caption{\label{pseumix}The $J^P=0^-$ resonance contributions in the decay
$J/\psi\to\gamma\phi\phi$ calculated in the model A which takes into account the mixing between the resonances $(X_1,X_2,X_3)\equiv[\eta(2225),\eta(2100),X(2500)]$.}\end{figure}
%--------------------------------------------------------------------------------
%--------------------------------------------------------------------------------
\begin{figure}
\includegraphics[width=12cm]{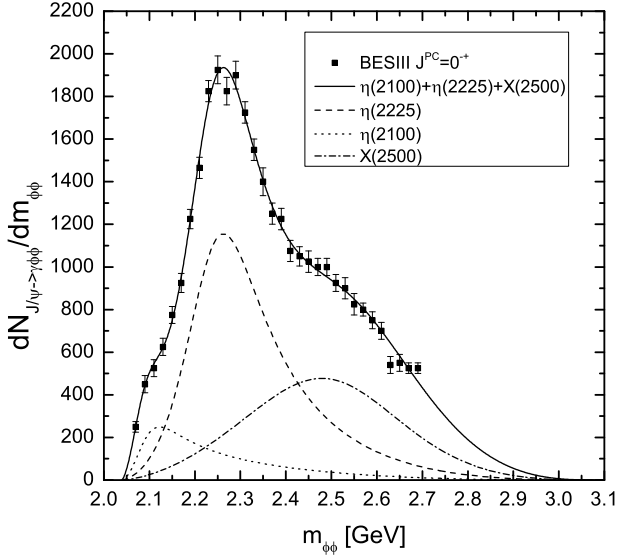}
\caption{\label{pseunomix}The same as in Fig.~\ref{pseumix}, but calculated in the model B which neglects the mixing of the $\eta(2225),\eta(2100),X(2500)$ resonances.} \end{figure}
%--------------------------------------------------------------------------------

The results of analysis of the $J^P=0^-$ partial wave contributions in the models A and B are plotted in Figs.~\ref{pseumix} and \ref{pseunomix}, respectively. One can see that the total contribution looks the same in both models, but the components corresponding to the resonances $\eta(2225),\eta(2100),X(2500)$ and their interference are different. A comparable values of $\chi^2/n_{\rm d.o.f.}$ in the models A and B show that these models cannot be distinguished  with the present accuracy of the data. Taking into account the present feature of the fit we will make further treatment in subsection \ref{subsec3b} in the  model B neglecting the mixing in situation when more that one resonance is required for the description of the data.

\subsection{Scalar resonance contribution}\label{subsec3b}
~

\begin{figure}
\includegraphics[width=12cm]{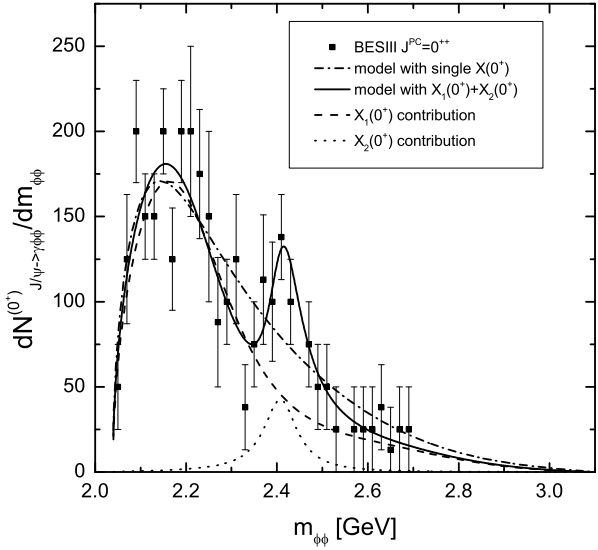}
\caption{\label{scal}The $J^P=0^+$ resonance contributions in the decay
$J/\psi\to\gamma\phi\phi$.}\end{figure}

The first attempt to describe the $J^{PC}=0^{++}$ resonance component of the spectrum uses the single resonance contribution given by the expression
\begin{eqnarray}\label{spec0pl1}
\frac{dN^{(0^+)}}{dm_{12}}&=&{\cal N}\frac{2m^2_{12}\Gamma_{J/\psi\to\gamma X(0^+)}\Gamma_{X(0^+)\to\phi\phi}}{\pi|D_{X(0^+)}|^2},
\end{eqnarray}where the $m_{12}$ dependencies of $\Gamma_{J/\psi\to\gamma X(0^+)}$ and   $\Gamma_{X(0^+)\to\phi\phi}$ are given by Eq.~(\ref{widthsc}), while $D_{X(0^+)}$ is given by Eq.~(\ref{prop}). The  parameters extracted from the fit are
\begin{eqnarray}\label{param0pl}
m_{X(0^+)}&=&2.381\pm0.018\mbox{ GeV},\nonumber\\
\Gamma^\prime_{X(0^+)}&=&0.001\pm0.025\mbox{ GeV},\nonumber\\
\sqrt{{\cal N}}g_1&=&361\pm9\mbox{ GeV}^{1/2},\nonumber\\
f_1&=&11.9\pm0.4\mbox{ GeV},\nonumber\\
f_2&=&0.9\pm0.6\mbox{ GeV}^{-1},\nonumber\\
B_{X(0^+)\to\phi\phi}&=&0.25\pm0.18,\nonumber\\
\chi^2/n_{\rm d.o.f.}&=&37.8/28\approx1.4.
\end{eqnarray}  Using  Eq.~(\ref{spec0pl1}) one obtains that
\begin{eqnarray*}
N^{(0^+)}&\equiv&{\cal N}\Gamma_{J/\psi}B_{J/\psi\to\gamma X(0^+)\to\gamma\phi\phi}=\nonumber\\&&\int_{2m_\phi}^{m_{J/\psi}}\frac{dN^{(0^+)}}{dm_{12}}dm_{12}=65\pm6.\end{eqnarray*}  One can see that the fit with the single scalar resonance is poor.
A better fit is obtained when adding the second $0^+$ resonance. To be specific, we neglect the mixing of the $X_1(0^+)$ and $X_2(0^+)$ analogously to the model B of the pseudoscalar contribution considered in subsection \ref{subsec3a}. The expression for the $J^{PC}=0^{++}$ resonance component of the spectrum  averaged over $\lambda_{J/\psi}=\pm1$ to be fitted is taken in the form
\begin{eqnarray}\label{spec0pl2}
\frac{dN^{(0^+)}}{dm_{12}}&=&\frac{{\cal N}}{(2\pi)^3\times12m^2_{J/\psi}}|{\bm k}||{\bm k}^\ast_1|\left(2|A_0|^2+\right.\nonumber\\&&\left.|A_0+A_2|^2\right),
\end{eqnarray}where ${\cal N}$ is the same unknown overall normalization factor as in Eq.~(\ref{spec0mi}). The amplitudes $A_0$ and $A_2$ are constructed using Eqs.~(\ref{fLSsc}), (\ref{widthsc}), and (\ref{spec0pl1}) and look as follows:
\begin{eqnarray}\label{A02}
A_0&=&\frac{g_{11}f_{001}}{D_{X_1(0^+)}}+\frac{g_{12}f_{002}}{D_{X_2(0^+)}},\nonumber\\
A_2&=&\frac{g_{11}f_{221}}{D_{X_1(0^+)}}+\frac{g_{12}f_{222}}{D_{X_2(0^+)}}.
\end{eqnarray}Here, the third index $i=1,2$ in $f_{00i}$ and $f_{22i}$, $i=1,2$, is introduced to designate the $X_i(0^+)$ contribution and are looking the same as in Eq.~(\ref{fLSsc}). In total, there are 10 fitted parameters: $m_{X_i}$, $\Gamma^\prime_{X_i}$, $g_{1i}$, $f_{1i}$, and $f_{2i}$; $i=1,2$. Recall that $\Gamma^\prime_{X_i}$  takes into  other possible decay modes besides the $\phi\phi$ one. One obtains the following  set of parameters accompanied by the relevant branching fractions:
\begin{eqnarray}\label{param0pl2}
m_{X_1(0^+)}&=&2.190\pm0.009\mbox{ GeV},\nonumber\\
\Gamma^\prime_{X_1(0^+)}&=&0.00\pm0.01\mbox{ GeV},\nonumber\\
\sqrt{{\cal N}}g_{11}&=&191\pm5\mbox{ GeV}^{1/2},\nonumber\\
f_{11}&=&8.5\pm0.3\mbox{ GeV},\nonumber\\
f_{21}&=&-6.9\pm1.2\mbox{ GeV}^{-1},\nonumber\\
B_{X_1(0^+)\to\phi\phi}&=&0.70\pm0.04,\nonumber\\
m_{X_2(0^+)}&=&2.409\pm0.010\mbox{ GeV},\nonumber\\
\Gamma^\prime_{X_2(0^+)}&=&0.003\pm0.021\mbox{ GeV},\nonumber\\
\sqrt{{\cal N}}g_{12}&=&-60\pm10\mbox{ GeV}^{1/2},\nonumber\\
f_{12}&=&-3.7\pm0.3\mbox{ GeV},\nonumber\\
f_{22}&=&1.4\pm0.7\mbox{ GeV}^{-1},\nonumber\\
B_{X_2(0^+)\to\phi\phi}&=&0.86\pm0.19,\nonumber\\
\chi^2/n_{d.o.f.}&=&19.7/23\approx0.9.
\end{eqnarray}For quantities characterizing the decay chain $J/\psi\to\gamma\phi\phi$ one obtains
\begin{eqnarray}\label{Nsc}
N^{(0^+)}&\equiv&{\cal N}\Gamma_{J/\psi}B_{J/\psi\to\gamma (X_1(0^+)+X_2(0^+))\to\gamma\phi\phi}=\nonumber\\&&63\pm5,
\end{eqnarray} and
 $${\cal N}\Gamma_{J/\psi}B_{J/\psi\to\gamma X_1(0^+)\to\gamma\phi\phi}=52\pm4,$$ $${\cal N}\Gamma_{J/\psi}B_{J/\psi\to\gamma X_2(0^+)\to\gamma\phi\phi}=5\pm2,$$ and the interference is $I=6\pm7$. The results of both fits described in this subsection are presented in Fig.~\ref{scal}.

%--------------------------------------------------------------------------------
%--------------------------------------------------------------------------------

\subsection{Tensor resonance contribution}
~

\begin{figure}
\includegraphics[width=12cm]{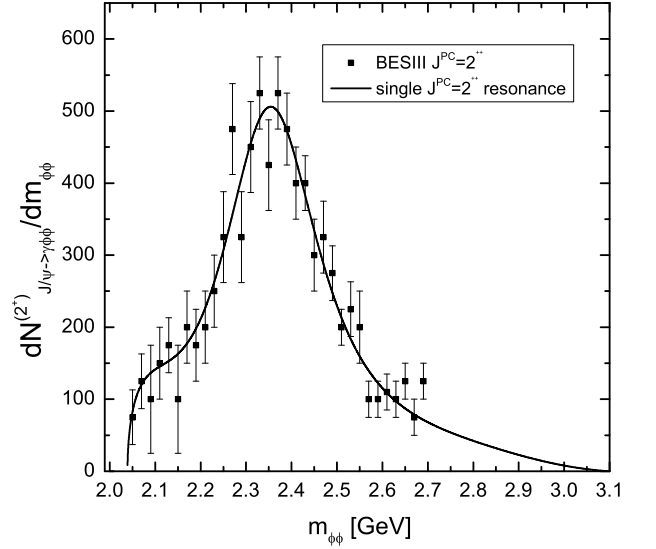}
\caption{\label{tens}The $J^P=2^+$ resonance contributions in the decay
$J/\psi\to\gamma\phi\phi$.}\end{figure}

In this case, one finds from Eq.~(\ref{Gampsi2}) that
\begin{eqnarray}\label{X2gamann}
\left\langle\Gamma_{J/\psi\to\gamma X(2^+)}\right\rangle_{\lambda_{J/\psi}=\pm1}&=&\frac{|{\bm k}|}{72\pi m^2_{J/\psi}}
\left[4g^2_{02}+\right.\nonumber\\&&\left.7g^2_{20}-4g_{02}g_{20}+\right.\nonumber\\&&\left.3(g_{12}+g_{20})^2\right].
\end{eqnarray}Note that the fitted parameters of the single $2^+$ contribution are $m_{X(2^+)}$, $\Gamma^\prime_{X(2^+)}$, $c_{1,2,3}$, and $g_{1,2,3,4}$, in total nine free parameters. Taking into account three $2^+$ contributions, as is made in Ref.~\cite{BES16}, requires 27 free parameters. So, having in mind a limited statistics of the data, we try to describe the $2^+$ component with the single tensor resonance with the help of parametrization
\begin{eqnarray}\label{spectens}
\frac{dN^{(2^+)}}{dm_{12}}&=&{\cal N}\frac{2m^2_{12}\left\langle\Gamma_{J/\psi\to\gamma X(2^+)}\right\rangle_{\lambda_{J/\psi}=\pm1}}{\pi|D_{X(2^+)}|^2}\times\nonumber\\&&\Gamma_{X(2^+)\to\phi\phi},
\end{eqnarray}where all necessary expressions are given by Eqs.~ (\ref{GamX2}), (\ref{prop}), and (\ref{X2gamann}). Surprisingly, but a rather good fit is obtained with the following set of parameters:
\begin{eqnarray}\label{param2pl}
m_{X(2^+)}&=&2.621\pm0.012\mbox{ GeV},\nonumber\\
\Gamma^\prime_{X(2^+)}&=&0.005\pm0.018\mbox{ GeV},\nonumber\\
\sqrt{\cal N}c_1&=&110\pm50\mbox{ GeV}^{-3/2},\nonumber\\
\sqrt{\cal N}c_2&=&2560\pm60\mbox{ GeV}^{-3/2},\nonumber\\
\sqrt{\cal N}c_3&=&-480\pm15\mbox{ GeV}^{1/2},\nonumber\\
g_1&=&-11.0\pm0.6\mbox{ GeV},\nonumber\\
g_2&=&25.0\pm1.6\mbox{ GeV}^{-1},\nonumber\\
g_3&=&-32.0\pm1.5\mbox{ GeV}^{-3},\nonumber\\
g_4&=&27.0\pm0.5\mbox{ GeV}^{-1},\nonumber\\
B_{X(2^+)\to\phi\phi}&=&0.21\pm0.01,\nonumber\\
\chi^2/n_{d.o.f.}&=&19.7/24\approx0.8.
\end{eqnarray}For quantities characterizing the decay chain $J/\psi\to\gamma X(2^+)\to\gamma\phi\phi$ one obtains
\begin{eqnarray}\label{Ntens}
N^{(2^+)}&\equiv&{\cal N}\Gamma_{J/\psi}B_{J/\psi\to\gamma X(2^+)\to\gamma\phi\phi}=\nonumber\\&&172\pm12.\end{eqnarray}
The contribution of the $2^{++}$ resonance to the $J/\psi\to\gamma\phi\phi$ spectrum evaluated with these parameters is shown in Fig.~\ref{tens}.
%--------------------------------------------------------------------------------
%--------------------------------------------------------------------------------
One should emphasize that the convergence to the minimal $\chi^2$ with the above parameters is very slow. In all appearance, this is due to the complicated dynamics of the $2^{++}$ partial wave contribution demanding nine free parameters entered in the nontrivial combinations corresponding to the given spin and orbital angular momentum of the $\gamma X(2^+)$ and $\phi\phi$ systems. [See Eqs.~(\ref{gLS}) and (\ref{fLS})]. So, in view of a limited accuracy of the present data, it seems to be prematurely to take into account three tensor resonances in the full dynamical form to describe the above contribution.

\begin{figure}
\includegraphics[width=12cm]{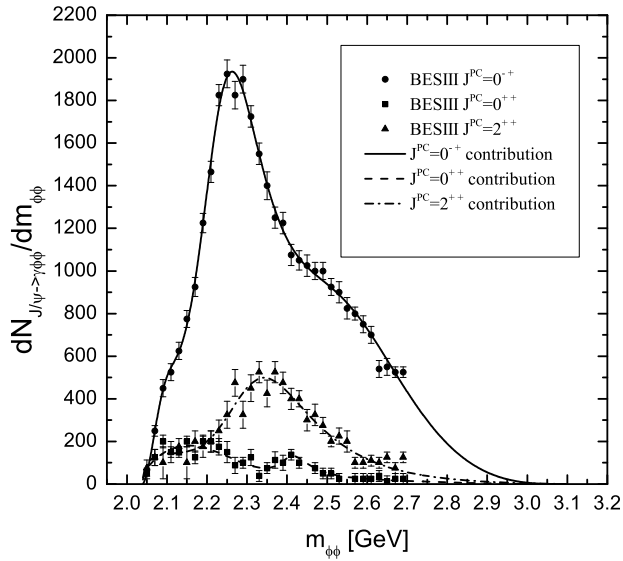}
\caption{\label{all}The fitted pseudoscalar, scalar, and tensor resonance contributions  in the $\phi\phi$ spectrum of the decay
$J/\psi\to\gamma\phi\phi$.}\end{figure}

The results of the fits of the considered resonance contributions to the $\phi\phi$ mass spectrum of the decay $J/\psi\to\gamma\phi\phi$ are summarized in Fig.~\ref{all}.
%--------------------------------------------------------------------------------
%--------------------------------------------------------------------------------

\section{Discussion and conclusion}\label{sec4}
~

Let us compare the resonance parameters such as masses and widths found in the present work, with the values given in Refs.~\cite{BES16} and \cite{PDG}. To this end, one should have in mind the following. The compatible with zero values of the widths $\Gamma^\prime_{X(J^{PC})}$ given in (\ref{param0pl}), (\ref{param0pl2}), and (\ref{param2pl}) refer to the contributions of the final states other than the $\phi\phi$. The latter  is taken into account explicitly, with the energy dependence of the contributions of various partial waves in the $J/\psi\gamma X(J^{PC})$ and $X(J^{PC})\phi\phi$ vertices. In the meantime, the parameters cited in Refs.~\cite{BES16} and \cite{PDG} were obtained in the fixed width approximation similar to Eq.~(\ref{simpleBW}). Hence the correct comparison of the results of the present work with the above references  requires the evaluation of the effective resonance peak positions and widths. A rough estimate can be obtained upon neglecting the resonance peak distortion due to the effects of the phase space volume. This can be made with help of Figs.~\ref{pseumix}, \ref{pseunomix}, \ref{scal}, and \ref{tens} by evaluating the width at the half of height of the resonance peaks. In the case  of the pseudoscalar resonances (in both models A and B of subsection \ref{subsec3a}) one finds the peak positions $m_{X_1(0^-)}\equiv m_{\eta(2250)}\approx2260$ MeV, $m_{X_2(0^-)}\equiv m_{\eta(2100)}\approx2120$ MeV, and $m_{X_3(0^-)}\equiv m_{\eta(2500)}\approx2480$ MeV while the effective widths are $\Gamma_{X_1(0^-)}\equiv \Gamma_{\eta(2250)}\approx220$ MeV, $\Gamma_{X_2(0^-)}\equiv \Gamma_{\eta(2100)}\approx210$ MeV, and $\Gamma_{X_3(0^-)}\equiv \Gamma_{\eta(2500)}\approx400$ MeV. Within one or  two magnitudes of the experimental uncertainty they agree with the values given in Ref.~\cite{BES16}. When fitting the scalar resonance contribution, the first one designated  here as $X_1(0^+)$, has the effective peak characteristics which, within the experimental accuracy, agree with those of the resonance $f_0(2100)$ observed in Ref.~\cite{BES16}. The second one, $X_2(0^+)$, included here to achieve the better description of the data, is new. However, taking into account rather large experimental error bars  in this sector, see Fig.~\ref{scal}, the latter conclusion should be treated as preliminary. The data with improved statistics could resolve the issue. The effective characteristics of the tensor resonance obtained here agree with those of $f_2(2340)$  cited in Ref.~\cite{PDG}.

\begin{figure}
\includegraphics[width=12cm]{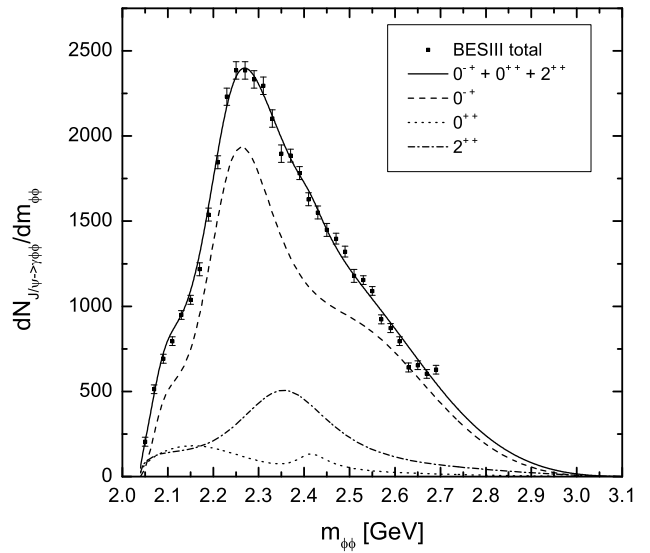}
\caption{\label{sum}Sum of fitted resonance contributions  to the
$J/\psi\to\gamma\phi\phi$ decay spectrum, together with the specific $J^{PC}$ ones.}\end{figure}

Let us check the consistency of the fits. First, one can evaluate the sum
\begin{equation}\label{totspec}
\frac{dN}{dm_{12}}=\frac{dN^{(0^-)}}{dm_{12}}+\frac{dN^{(0^+)}}{dm_{12}}+\frac{dN^{(2^+)}}{dm_{12}}\end{equation} and plot the result to compare  with the data \cite{BES16}. The results are shown in Fig.~\ref{sum}. For comparison, also shown are the curves corresponding to the specific $J^{PC}$ contribution. One can see that the contributions with different quantum numbers $J^P$  add incoherently as they should. The formal reason is briefly explained in Appendix.
%----------------------------------------------------------------------------
%--------------------------------------------------------------------------------

\begin{figure}
\includegraphics[width=12cm]{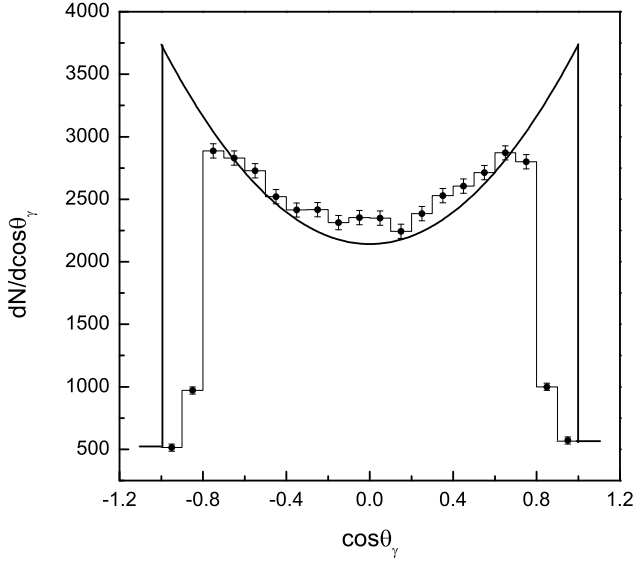}
\caption{\label{angdist}Angular distribution of photons in the decay
$J/\psi\to\gamma\phi\phi$. Histogram -- BESIII data \cite{BES16}. The solid curve -- evaluation with the resonance parameters found  from the fits.} \end{figure}

Second, using expressions for the amplitudes, one can obtain the expression for  the angular distribution of final photons in the decay $J/\psi\to\gamma\phi\phi$. One has
\begin{eqnarray}\label{angular}
\frac{dN}{d\cos\theta_\gamma}&=&\frac{3}{8}(1+\cos^2\theta_\gamma)\left[N^{(0^-)}+N^{(0^+)}\right]+\nonumber\\&&N^{(2^+)}_1+
N^{(2^+)}_2\cos^2\theta_\gamma,
\end{eqnarray}where the central values $N^{(0^-)}=708$ and $N^{(0^+)}=63$ are given in the Table \ref{tab1} and by Eq.~(\ref{Nsc}), respectively, and
\begin{widetext}
\begin{eqnarray}\label{N12}
N^{(2^+)}_{1,2}&=&\frac{{\cal N}}{32\pi m^2_{J/\psi}}\int_{2m_\phi}^{m_{J/\psi}}dm_{12}\frac{m^2_{12}\Gamma_{X(2^+)\to\phi\phi}(m_{12})}{\pi\left|D_{X(2^+)}\right|^2}|{\bm k}|\left[|g_{02}|^2+\frac{3}{2}|g_{20}|^2-\frac{1}{3}|g_{02}+g_{20}|^2\pm|g_{12}+g_{20}|^2\right].
\end{eqnarray}
\end{widetext}With the set of parameters (\ref{param2pl}) one finds that the central values are $N^{(2^+)}_1=89$ and $N^{(2^+)}_2=-8$. Multiplying Eq.~(\ref{angular}) by the normalization factor $N^{-1}$ where $$N=N^{(0^-)}+N^{(0^+)}+2N^{(2^+)}_1+\frac{2}{3}N^{(2^+)}_2,$$
one obtains the angular distribution normalized to unity. Multiplying the obtained expression by the area under the experimental histogram in Fig.~\ref{angdist} one gets the curve shown with the solid line in Fig.~\ref{angdist}.
%--------------------------------------------------------------------------------
%--------------------------------------------------------------------------------
So, one can see that without separate fitting of the total spectrum Eq.~(\ref{totspec}) and the photon angular distribution Eq.~(\ref{angular}), their evaluated magnitudes agree with the data. The above evaluations  support the consistency of the fits of the separate resonance contributions.

Some concluding remarks are in order. The dynamical analysis of the resonance contributions to the $J/\psi\to\gamma X\to\gamma\phi\phi$ decay amplitude is presented based on the effective amplitudes of the transitions $J/\psi\to\gamma X$ and $X\to\phi\phi$. The $X$-resonances with the quantum numbers $J^{PC}=0^{-+}$, $0^{++}$, and $2^{++}$ are taken into account to describe the $\phi\phi$ mass spectrum in the decay $J/\psi\to\gamma X(J^{PC})\to\gamma\phi\phi$ studied by BESIII collaboration \cite{BES16}. Two models, with and without  mixing of three $X(0^{-+})$ resonances, are considered when fitting the pseudoscalar component of the spectrum. It is shown that both above models give satisfactory description of the data, hence one cannot distinguish between them with the present accuracy of the data. The scalar component of the $\phi\phi$ spectrum is better described in the model with two scalar resonances. Surprisingly, the tensor component requires only one resonance, because the non-trivial  behaviour shown in Fig.~\ref{tens} at the left shoulder of the resonance peak is due to the dependence on the $\phi\phi$ invariant mass of the contributions with given spin and orbital angular momentum in the $X(2^{++})\to\phi\phi$ vertex. Masses and effective coupling constants parametrizing invariant amplitudes are extracted from the fits and used for evaluation of branching fractions. The consistency of the fits is supported by the evaluation of the incoherent sum of the $0^{-+}$, $0^{++}$, and $2^{++}$ resonance contributions to the $\phi\phi$ mass spectrum of the reaction $J/\psi\to\gamma\phi\phi$ and of the angular distribution of the final photons.  Their  calculated magnitudes are shown to agree with the data Ref.~\cite{BES16}.

The work was supported by the program of fundamental scientific researches of the Siberian Branch of the Russian Academy of Sciences no. II.15.1., project no. 0314-2016-0021.

\appendix*
\section{}
\label{app}
~

It is well known that the helicity selection rule $\lambda_{J/\psi}=\lambda_\gamma-\lambda_X$ leaves only one independent helicity amplitude for zero spin of $X$ resonance and three helicity amplitudes in case of the spin two resonance, in the $J/\psi\to\gamma X(J^P)$ transition amplitude \cite{zou}. So let us give the expressions for the spin structure of the amplitudes $J/\psi\to\gamma X(J^P)\to\gamma\phi\phi$ in terms of the helicity amplitudes $M^{(J^P)}_{\lambda_{J/\psi},\lambda_\gamma,\lambda_X}$ of the decay $J/\psi\to\gamma X(J^P)$. They are necessary for checking the fact that the distinct $J^P$ contributions to the $J/\psi\to\gamma\phi\phi$ mass spectrum do not interfere. For the sake of simplicity,  the single intermediate $X(J^P)$ resonance will be assumed. The helicity amplitudes of the decay $J/\psi\to\gamma X(J^P)$ are calculated from Eqs.~(\ref{ampga0mi}), (\ref{Mga0pl}), and (\ref{Mga2pl}). The expressions of the above amplitudes $M^{(J^P)}_{\lambda_{J/\psi},\lambda_\gamma,\lambda_X}$ are the following:
\begin{eqnarray}\label{inassign}
M^{(0^-)}_{1,1,0}&=&-ig_{J/\psi\gamma X(0^-)}m_{J/\psi}|{\bm k}|,\nonumber\\
M^{(0^+)}_{1,1,0}&=&g_1,\nonumber\\
M^{(2^+)}_{1,1,0}&=&\frac{1}{\sqrt{6}}(2g_{02}-g_{20}),\nonumber\\
M^{(2^+)}_{0,1,1}&=&-\frac{1}{\sqrt{2}}(g_{12}+g_{20}),\nonumber\\
                                         M^{(2^+)}_{-1,1,2}&=&-g_{20},\end{eqnarray}where $g_{02}$, $g_{20}$, and $g_{12}$ are given by Eq.~(\ref{gLS}).

Then one gets the following expressions for the $J/\psi\to\gamma X(J^P)\to\gamma\phi\phi$ transition amplitudes:
\begin{widetext}
\begin{eqnarray}\label{helamps}
M_{J/\psi\to\gamma X(0^-)\to\gamma\phi\phi}&=&\frac{iM^{(0^-)}_{1,1,0}f^{(0^-)}_{11}}{D_{X(0^-)}(m^2_{12})}({\bm n}\cdot[{\bm\xi}\times{\bm e}])e_{abc}n_{1c}\xi_{1a}\xi_{2b},\nonumber\\
M_{J/\psi\to\gamma X(0^+)\to\gamma\phi\phi}&=&\frac{M^{(0^+)}_{1,1,0}}{D_{X(0^+)}(m^2_{12})}({\bm\xi}\cdot{\bm e})\left[f^{(0^+)}_{00}\delta_{ab}+f^{(0^+)}_{22}n_{1a}n_{1b}\right]\xi_{1a}\xi_{2b},\nonumber\\
M_{J/\psi\to\gamma X(2^+)\to\gamma\phi\phi}&=&\left\{\frac{1}{2}\left(\sqrt{6}M^{(2^+)}_{1,1,0}-M^{(2^+)}_{-1,1,2}\right)({\bm\xi}\cdot{\bm e})n_in_j-M^{(2^+)}_{-1,1,2}\xi_{\bot i}e_j-\sqrt{2}M^{(2^+)}_{0,1,1}({\bm\xi}\cdot{\bm n})e_in_j\right\}\times\nonumber\\&&\left[f^{(2^+)}_{20}\delta_{ka}\delta_{lb}+f^{(2^+)}_{02}
\delta_{ab}n_{1k}n_{1l}+f^{(2^+)}_{22}(n_{1a}\delta_{kb}+n_{1b}\delta_{ka})n_{1l}+f^{(2^+)}_{24}n_{1a}n_{1b}n_{1k}n_{1l}\right]\times\nonumber\\&&
\frac{\Pi_{ij,kl}}{D_{X(2^+)}(m^2_{12})}\times\xi_{1a}\xi_{2b}.
\end{eqnarray}
\end{widetext}Here, the amplitudes $f^{(J^P)}_{SL}$ with the spin orbital momentum assignment $(S,L)$ of the $\phi\phi$ state are expressed through the coupling constants characterizing the invariant amplitudes of the decay $X(J^P)\to\phi\phi$. In the case of the $0^-$ state one has
\begin{eqnarray}\label{finassign}
f^{(0^-)}_{11}&=&g_{X(0^-)\phi\phi}m_{12}|{\bm k}^\ast_1|,
\end{eqnarray}while in the case of the $0^+$ state they are given by Eq.~(\ref{fLSsc}), and in the case of the $2^+$ state $f^{(2^+)}_{SL}\equiv f_{SL}$ are given by Eq.~(\ref{fLS}).

The direct calculation of the final probability distribution shows that, when summed over polarizations of the final $\phi$ mesons but keeping fixed their direction of motion, the vanishing are the interference terms of the contributions with opposite parities.  In turn, the
$(0^+-2^+)$  interference term being  proportional to $n_{1k}n_{1l}-\delta_{kl}/3$,  vanishes after the integration over phase space of the $\phi$ mesons because the averaging over the $\phi$ meson direction of motion results in the relation
\begin{equation}\label{nn}
\left\langle n_{1i}n_{1j}\right\rangle=\frac{1}{3}\delta_{ij}.
\end{equation}Hence, all the considered $J^{PC}=0^{-+}$, $0^{++}$, and $2^{++}$ contributions to the $\phi\phi$ mass spectrum and to the photon angular distribution in the decay $J/\psi\to\gamma\phi\phi$ do not interfere and are added incoherently. This conclusion agrees with the experimentally verified fact. See Table II in Ref.~\cite{BES16}


\begin{thebibliography}{99}
\bibitem{fn1}
To some extent, the mixing of resonances considered below in Eq.~(\ref{Amix}), can mimic the effect of the final state interaction (FSI). The fact is that, in view of the unitary relation, in the three-step transition $X_i\to\phi\phi\to X_j\to\phi\phi$ due to the mixing, the last two  steps $\phi\phi\to X_j\to\phi\phi$  represent  the s-channel contribution of the $\phi\phi\to\phi\phi$ reaction amplitude. From the duality point of view, the sum of all s-channel resonance contributions over large but finite energy interval should coincide with the sum of all t-channel exchange contributions. So, to take into account FSI in the $X_i\to\phi\phi$ decay, one should insert the full $\phi\phi\to\phi\phi$ amplitude calculated either as the sum of all s- or t-channel exchanges, not both. However, the usual practice (adopted in the present work) is to take into account a few resonances for the process proceeding in the restricted energy interval. In such a case the $X_i\to\phi\phi\to X_j\to\phi\phi$ transition amplitude has the s-channel pole corresponding to the resonance $X_j$, and is attributed to the mixing in Eq.~(\ref{Amix}). The diagram with the t-channel exchange gives rise to the imaginary part of the $X_{i,j}\phi\phi$ coupling constant. It is just this contribution which is neglected in the present work.
\end{thebibliography}
\end{document}